# Azimuthal imaging of rock fractures by incorporating single borehole radar and optical data

**Jian Shen** [1,2]**, Liu Liu** [2*]**, Shaojun Li** [2]**, Zhenming Shi** [1]**, Yiteng Wang** [2]**, Ming Peng** [1]**, and Minzong Zheng** [2]

1 Department of Geotechnical Engineering, Key Laboratory of Geotechnical and Underground Engineering of Ministry of Education, Tongji University, Shanghai, 200092, China

2 State Key Laboratory of Geomechanics and Geotechnical Engineering, Institute of Rock and Soil Mechanics, Chinese Academy of Sciences, Wuhan, 430071, China.

\* Corresponding author: Liu Liu. E-mail: liuliu@mail.whrsm.ac.cn

**Abstract**

Single borehole radar detection suffers from azimuthal ambiguity, while borehole optical tests only provide information about the borehole wall. These limitations prevent either detection method from revealing the complete spatial patterns of rock fractures on their own. In this paper, we address these challenges by proposing a joint imaging method that combines the advantages of both borehole detection methods. Geological azimuthal parameters are extracted from optical images by fitting the fracture curves to sinusoidal functions. A 2D Kirchhoff time migration is then implemented using radar common offset gather. Up-dip and down-dip events are separated by the f-k transform or z-s transform, depending on their geometric relation. The complete fracture planes, including trend, dip angle, gap width, and extension length, are finally reconstructed in 3D space by mapping the migration profile using azimuthal information from optical images. The method is proven reliable and high-resolution through both numerical tests and real field data.

***Keywords***: Single borehole radar; Borehole optical imaging; Rock fracture; Azimuthal imaging; Migration

## 1. Introduction

Rock mass exhibits strong discontinuity since it comprises numerous weak structural planes (such as joints,

faults, fractures, etc.) (Yu *et al*. 2023). These structural planes affect the overall quality and strength of the excavated rock mass (Wu *et al*. 2022). Trend, dip angle, gap width and extension length are four critical parameters for describing a fracture plane. Borehole drilling and loggings are most common practice to investigate these geometry parameters of factures (Liu *et al*. 2024a, Liu *et al*. 2024b).

Single borehole radar (simplified as borehole radar in the following paragraphs) is a geophysical exploration equipment developed especially for narrow borehole environment (Liu *et al*. 2004, Liu *et al*. 2012). The common offset gather is obtained by moving the borehole dipole radar antenna along the borehole axis. Generally, applying migration algorithms (Huo *et al*. 2020, Huo *et al*. 2019) is a reliable way to image the fractures and retrieve their extension lengths. Nevertheless, due to the omnidirectional radiation pattern of dipole antennas, radar data lacks azimuthal information. Although several efforts have been made to alleviate such disadvantages by direction of arrival (DOA) analysis (Liu *et al*. 2019, Sato and Takayama 2007, Takayama and Sato 2007), the intrinsic azimuthal ambiguity of radar migration profiles still exists.

Borehole optical imaging is another logging method which captures the pictures of borehole wall and maps them into planar images. It is assumed that projection trajectories of an intersecting-borehole fracture plane in a 2D image is ideally a sinusoidal curve (Li *et al*. 2019). By segmenting the sinusoidal curves automatically from the planar optical image and fitting them to standard sinusoidal functions (Wang *et al*. 2022), the trends, dip angles and gap widths of fractures can be successfully extracted. However, the information of fractures outside the borehole is completely in vacancy as far as the optical imaging is concerned.

In this paper, we propose an integrated borehole geophysical detection method which combines the borehole optical imaging with borehole radar migration imaging to reconstruct the rock fracture planes in 3D model space. Firstly, through analysing the 2D borehole optical image, the geological parameters like trend, dip angle and gap width can be extracted. Then, the migration profile of borehole radar data is generated by implementing Kirchhoff migration. The two branches of fracture planes at the opposite side of borehole are separated by f-k transform or z-s transform depending on fracture trends. The 2D migration profile can be easily mapped to 3D space based on the azimuthal information provided by the optical image. The process and performance of this strategy is illustrated using two typical synthetic borehole models at first and then a real field data. Some discussion on the limitation of this method is provided at last.

## 2. Methods

In this section, we further introduce the motivation of this research and how the problem of non-uniqueness during borehole radar and optical test is raised. Then, we elaborate on the proposed imaging method and reconstruct the fracture planes in 3D space.

2.1 Motivation

The schematic of implementing digital panoramic borehole imaging and borehole radar detection is demonstrated in Fig. 1. Both the optical probe and radar probe are released from the top of a borehole and subsequently moved deeper along the axis of borehole through a data transferring cable.

As shown in Fig. 1(a), the panoramic camera at the end of the probe captures the 360° view of borehole wall (highlighted by yellow region). A 2D planar image (Fig. 1(c)) is mapped from the raw panoramic photo. The vertical direction indicates the depth. The four symbols N, E, S, and W in the horizontal direction represent the four directions of magnetic north, east, south, and west, respectively. The intersecting-borehole fracture plane manifests itself as a sinusoidal curve in the image. By fitting it to a standard sinusoidal function, the geological parameters, for example, the trend $\varphi$, dip angle $\theta$, and fracture gap width $d$ can be calculated.

On the other hand, the borehole radar adopts the same detection strategy as GPR in which the electromagnetic waves are transmitted and received at a constant space interval. The reflection from the fracture plane creates a 'V' pattern in the common offset gather (Fig. 1(d)). The two branches, referred to as up-dip and down-dip, represent the up-propagating wave and down-propagating wave respectively. The vertex of the 'V' pattern indicates the intersection point of fracture and borehole. The angle $\alpha$ between the branch of 'V' and the horizontal direction is not strictly equal to the dip angle as long as the fracture plane is not parallel to the borehole. Only through wave field migration can the reflection energy be converged to its reflector, which leads to the recovery of the true dip angle and the extension length towards the outside of the borehole. However, the azimuthal information is still missing due to the omnidirectional radiation pattern of the borehole dipole radar antenna.

The question is clearly explained by the following two numerical borehole models. The corresponding raw detection data in Fig. 1 exhibits a vivid contradiction. Model 1 in Fig. 1(e) contains two fracture planes with the same dip angle, trend, and gap width but different extension lengths while Model 2 in Fig. 1(f) encompasses two fractures with the same dip angle and extension length but different trends and gap widths. Fig. 2(g), (i) and

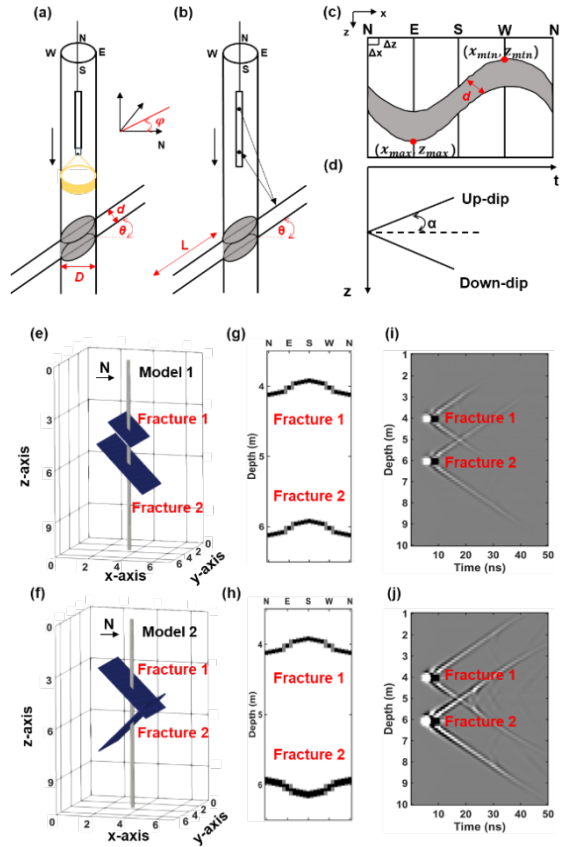

Fig 1. The schematic of implementing borehole optical and radar detection with two typical numerical borehole models and their corresponding detection results. (a) Borehole panoramic optical imaging. (b) Borehole radar. (c) 2D planar optical image. (d) 'V' shape pattern of fracture in radar record. (e and f) Borehole model 1 and 2. (g and h) Pseudo borehole optical images of model 1 and 2. (i and j) Synthetic radar common offset gathers of model 1 and 2.

(h), (j) respectively demonstrate their corresponding pseudo-optical images and synthetic radar profiles. As shown in Fig. 1(g), it is impossible to tell the difference between the two fractures of Model 1 merely from the optical image. Nevertheless, the radar profile (Fig. 1(i)) suggests that the extension length of the first fracture is shorter. In contrast, the two fractures of Model 2 have the same patterns in the radar profile (Fig. 1(h)) except the reflection energy of the second fracture is slightly stronger because of its larger gap width. But they are completely different in the optical borehole image. To correctly image the shape and attitude of fracture planes, it is undoubted that the optical data and radar data should be analysed jointly and complement each other.

Thus, whether borehole optical imaging or borehole radar imaging alone is not capable of revealing the full morphology of fracture planes in the 3D space. The motivation of this study arose from the practical requirement for an effective and convenient imaging method that combines the advantages of optical detection and radar detection in terms of delineating intersecting borehole fracture planes.

2.2 Imaging Method

The whole flowchart of the proposed method is demonstrated in Fig. 2. The geological geometry parameters of the fracture plane are retrieved through the optical image while the fracture length is acquired through a 2D borehole radar migration profile. The fracture planes are finally reconstructed in 3D space.

1) Geological geometry parameter extraction: The curve of a fracture plane in the optical image can be fit to a standard sinusoidal function as follows (Wang *et al.* 2022):

$$z(x) = z_0 - M sin(\omega x + \beta). \tag{1}$$

in which $x, z$ denote the horizontal and vertical coordinates and are discretized by image pixel resolution in these two directions, namely $\Delta x, \Delta z$, $z_0$ denotes the vertical coordinate of the center of sinusoidal function on the image, $\omega$ denotes the angular frequency of sinusoidal function, $\beta$ denotes the initial phase of sinusoidal curve and $M$ denotes the maximum amplitude. The trend of fracture $\varphi$ can be calculated as follows:

$$\varphi = \frac{2 x_{max} \Delta x}{D}. \tag{2}$$

in which $x_{max}$ represents the horizontal coordinate of the point on sinusoidal curve with maximum z value and $D$ represents the diameter of borehole. The dip angle $\theta$ can be calculated as follows:

$$\theta = \arctan\left(\frac{(z_{max} - z_{min})\Delta z}{D}\right). \tag{3}$$

in which $z_{min}$ ($z_{max}$) represents the vertical coordinate of the point on sinusoidal curve with minimum(maximum) z value. The gap width can be expressed as follows:

$$d = \frac{\sum_x \left(z_{lower}(x) - z_{upper}(x)\right) \Delta x \Delta z \cos\theta}{\pi D}. \tag{4}$$

in which $z_{lower}(x)$ and $z_{upper}(x)$ represents the lower-bound and upper-bound of sinusoidal curve at each coordinate $x$.

2). Borehole radar migration: To converge the reflection energy to its reflector and retrieve the actual length of fracture plane, the Kirchhoff migration algorithm is adopted in this study. The core equation expressing the idea of Kirchhoff migration can be written as follows (Docherty 1991):

$$u(x, z, t = 0) = \frac{1}{2\pi} \int_{-\infty}^{+\infty} \left[\frac{\cos\theta}{r^2} u(0, z_r, \tau) + \frac{\cos\theta}{V} \frac{\partial}{\partial \tau} u(0, z_r, \tau)\right] dz_r. \tag{5}$$

where $u(x, z, t = 0)$ represents the migrated wave field $u(x, z)$ at initial time $t = 0$, $(x, z)$ is the coordinate along the radial direction with respect to borehole and along the axial direction and, $u(0, z_r, \tau)$ represents the recorded wave field by the receiver at $(x = 0, z = z_r)$, $r$ is the distance between the image point and the receiver point, $V$ is the velocity, $\tau = \frac{r}{V}$ denotes the time advance and $\theta$ denotes the incident angle. Since it is unfeasible to directly obtain velocity from the borehole radar profile, here we only implement the time migration with a constant velocity estimated based on rock lithology.

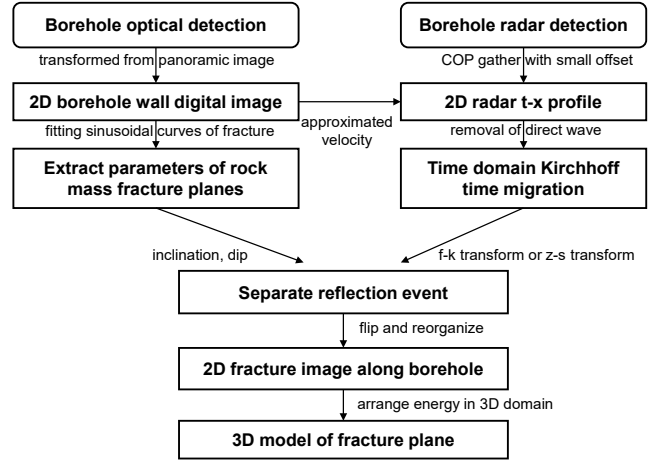

Fig 2. Flowchart for the joint borehole imaging method.

After migration, the up-dip and down-dip (Fig. 1(d)) event still remain the 'V' shape. To firstly reconstruct the fracture plane in 2D panel, these two events should be separated and one of them is flipped along the borehole axis. One of the most prominent steps is to select the appropriate filters to separate the up-dip and down-dip events correctly depending on their trend. When there is no intersection between an up-arm and a down-arm, a simple *f-k* transform is enough to separate these two events apart based on their respective negative and positive slopes. On the contrary, if an up-arm and a down-arm crosses in 3D space (Fig. 1(f)), they should remain together. Instead of *f-k* transform, we define a $\tau - p$ alike transform method called $z - s$ transform which is essentially a slant stacking along the depth axis of migration profile using different slopes *s* as equation 6:

$$m(z,s) = \int_0^{t_{max}} d(z = z_0 + st, t)dt. \qquad (6)$$

in which $d(z = z_0 + st, t)$ is the migration profile and $m(z, s)$ is the $z - s$ panel.

In the $z - s$ panel, the target up-dip and down-dip events can be both reserved by muting the unwanted energy. Then the $z - s$ panel will be inversely transformed back to the migration profile to meet the purpose of reconstructing the fractures in the space domain. Finally, the 2D migration profile can be mapped to 3D space by conforming with the trend and dip angle of each fracture.

## 3. Numerical Results

The fracture parameters of two numerical 3D models displayed in Fig. 1 are listed in Fig. 3 (i). The models are discretized by rectangular grids in 0.02*0.02*0.02 m spacing. The borehole diameter is set as 0.2 m. The pseudo-optical image is generated by mimicking the observation mode of a borehole panoramic camera and picking the grid points corresponding to the borehole wall. The geological parameters extracted from pseudo-optical images are also listed in Fig. 3(i) as 'observed' (trend, dip angle, and gap width). The ground true and calculated value is highly consistent and the residuals are most likely caused by model discretization. The synthetic radar profile is generated by performing forward calculation of Maxwell's equations using the finite difference time domain method in 3D space. The whole process is accomplished by utilizing the open-source software gprMax (Warren *et al*. 2016). The relative permittivity of the surrounding rock is set to be 6. Since both the borehole and fracture are assumed to be filled with air, their relative permittivity is set to be 1. The excitation source is a Ricker wavelet with a central frequency of 200 MHz. The whole simulated time is 50 ns and the temporal step is 0.04 ns.

Based on optical images, it is apparent that the fractures in Model 1 are parallel to each other. By applying the f-k filter to the migration profile, the two branches of the 'V' pattern are separated, as depicted in Fig. 3(a) and (b). One arm of the fracture is flipped and concatenated with the other arm to reconstruct the crossing-borehole fracture in the 2D migration profile (Fig. 3(c)). The 2D profile is subsequently mapped to 3D space based on the azimuthal information provided by the optical image (Fig. 3(g)). It should be pointed out that since only one dimension of a fracture plane is reflected in the migration profile, the extension length in the other dimension is assumed as 4 m long in this study.

However, the same process cannot be directly repeated in the case of Model 2 because the up-dip and down-dip event intersects. Therefore, the $z-s$ transform is performed instead. The strongest energy concentrates around the area that corresponds to the slope of +1 and -1 and the depth of 4 m and 6 m, which are identical to the dip angle of 45° and the crossing point between borehole and fracture planes. The migration profile after the appropriate filtering in $z-s$ panel and inverse $z-s$ transform is shown in Fig. 3(d). Only the branches of expected slopes are preserved despite that a certain degree of energy distortion and leakage happen around the endpoints of fracture traces because $z-s$ transform is not orthogonal. Fig. 3(e) shows the other half of the fracture which is also flipped to form a complete fracture image in the migration profile (Fig. 3(f)). Fig. 3(h) is the 3D presentation of Fig. 3(f).

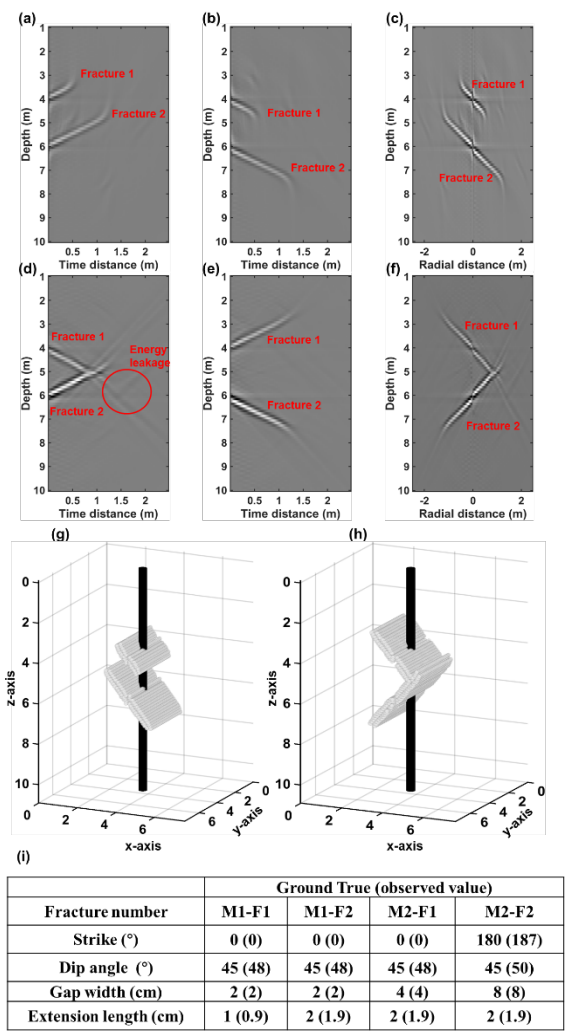

Fig 3. Separated migration profiles and 3D reconstructed fracture images of the numerical model. (a and b). The separated up-dip and down-dip event of Model 1. (c) The reconstructed 2D fracture migration profile of Model 1. (d and e). The separated up-dip and down-dip event of Model 2. (f) The reconstructed 2D fracture migration profile of Model 2. (g and h) The reconstructed 3D fracture images of Model 1 and Model 2. (i) The fracture parameters set and observed in two synthetic borehole models. The designator M1-F1 stands for Fracture 1 in Model 1 and others obey the same abbreviation rules.

## 4. Field data validation

In this section, we utilize a real field example from the prospecting project of Sujiapo phosphate ore to validate the proposed imaging method. Sujiapo phosphate mine is located in the territory of Lotus Town, Yuanan County, Yichang City, Hubei Province, with a burial depth of 300-950 m and an average thickness of 2.71 m. Its regional tectonic position is in the turning part of the north-east wing of the Huangling Dorsal Plateau. To analyse the initial rock stress characteristics of deep phosphate mines, the hydraulic fracturing ground stress measurements had been carried out. Prior to the in-situ stress, it was critical to determine the optimal testing points in the boreholes based on the distributions of rock fractures. Therefore, the borehole optical and radar detection was implemented. Fig. 4(a) demonstrates several segments of an entirely 17 m-long borehole optical image from one of the testing boreholes. The diameter of the borehole was 0.1 m. The single borehole radar being utilized in the test was developed by Wuhan Changshengmeian technology CO., LTD. It consisted of a surface computer, the transmitting and receiving antenna, the wire cable, the depth counter, and the charger. The system features a small diameter (36 mm) and near offset (30 cm). The electromagnetic waves with the central frequency of 100 MHz were transmitted and received every 0.5 m as moving along the borehole axis. Fig. 4(b) and Fig. 4(c) present the original borehole radar data and the one after the removal of the direct wave component. The counterparts for six fracture curves in the optical image can all be found at ease in the migration profile as shown in Fig. 4(d).

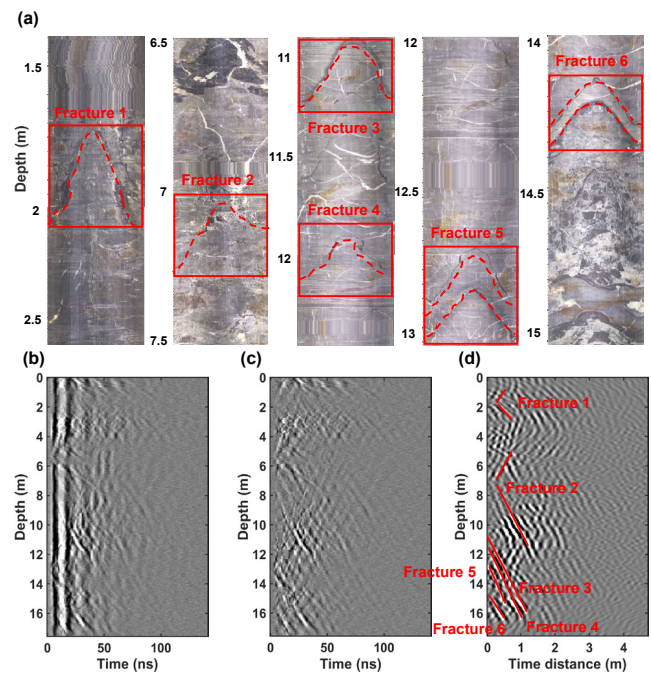

Fig 4. Field borehole optical data and radar data. (a) Five segments of 17 m borehole optical image. The red dashed lines denote the traces of fractures. (b) Raw radar record. (c) Radar record after removal of direct wave. (d) Migration profile of radar record. The reflection corresponding to fractures in (a) is marked by red lines.

The optical images suggest that the trends of all six fractures are approximately the same as being northbound, hence only an f-k filter is applied to separate the up-dip and down-dip events. The reconstructed 2D fracture and 3D fracture planes are demonstrated in Fig. 5(a) and (b) respectively. The attitude parameters and extension lengths of six fractures are all listed in Fig. 5(c). Most of the fractures have two halves of different lengths, some even have only one half (Fracture 4 and Fracture 6), which is probably caused by the fracture discontinuity at the intersecting point with the borehole. It is worth noticing that Fractures 3-6 are very close to each other but are still successfully distinguished in the radar migration profile with the aids from optical images. The field data profoundly proves that our method, by incorporating optical detection and radar detection, can alleviate the azimuthal ambiguity of the conventional borehole detection method and image the entire fracture plane spreads with high resolution.

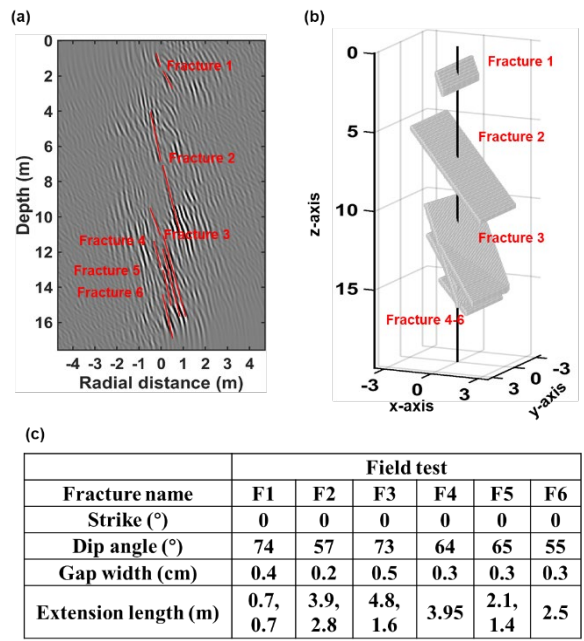

| Fracture name | F1 | F2 | F3 | F4 | F5 | F6 |
|---|---|---|---|---|---|---|
| Strike (°) | 0 | 0 | 0 | 0 | 0 | 0 |
| Dip angle (°) | 74 | 57 | 73 | 64 | 65 | 55 |
| Gap width (cm) | 0.4 | 0.2 | 0.5 | 0.3 | 0.3 | 0.3 |
| Extension length (m) | 0.7, 0.7 | 3.9, 2.8 | 4.8, 1.6 | 3.95 | 2.1, 1.4 | 2.5 |

Fig 5. Reconstructed 2D migration profile and 3D reconstructed fracture image of field data example. (a). Reconstructed 2D migration profile containing 6 fractures. (b) The reconstructed 3D fracture image. (c) The fracture parameters obtained from field data. The abbreviation 'F' stands for 'Fracture'

**5. Discussions and conclusions**

This study proposes a novel imaging method that combines single borehole radar and borehole optical data to quantitatively and azimuthally reconstruct intersecting borehole fracture planes in 3D space. Azimuthal parameters of rock fractures are extracted from optical images by fitting the fracture curves to sinusoidal functions. Subsequently, a 2D Kirchhoff time migration is implemented using radar common offset gather. The up-dip and down-dip events can be separated by f-k transform or z-s transform depending on their geometric relation. The complete fracture planes, including trend, dip angle, gap width, and extension length, are finally reconstructed in 3D space by mapping the migration profile using azimuthal information acquired from optical images. The proposed method has been verified on both synthetic data and real field data.

Some aspects still need further study. Firstly, only one dimension of a fracture plane can be precisely imaged, with the length in the other dimension set as default. However, some reflection events may indicate the edge of a fracture plane (Fig. 4(c), (d)), which could be utilized in future studies to infer the length in the second dimension. Additionally, the migration technique used in this study is time migration, which assumes the surrounding rock is homogeneous and the velocity is a pre-known constant. However, real-world cases are often more complex, which may affect the imaging quality of borehole radar. A better prior understanding of rock properties, obtained from laboratory tests on core samples, and depth migration could help address this issue. Furthermore, since the z-s transform is not a standard orthogonal transform, distortion, and leakage of migration energy occur, especially around the endpoints of the fracture image, after filtering in the z-s panel. As a result, the calculation of the extension lengths of fracture planes is affected, though often only to a minor degree.

**Funding**

This work was supported by the National Natural Science Foundation of China, Grant No., 42207211 and 42172296.

**Conflict of interest statement**

The authors declare that they have no conflicts of interest.